\documentclass[runningheads]{llncs}

\usepackage{blindtext}
\usepackage{multicol}
\usepackage{soul,color}
\usepackage[htt]{hyphenat}
\usepackage{booktabs} 
\usepackage{mathtools}
\usepackage{listings}
\usepackage{graphicx}
\usepackage{amssymb}
\usepackage{amsmath}
\usepackage{mathtools}
\usepackage{adjustbox}
\usepackage{float}
\usepackage{psfrag}
\usepackage[mathscr]{euscript}
\usepackage{acronym}  
\usepackage{booktabs}
\usepackage{graphicx}
\usepackage{amsmath}
\usepackage[algo2e,ruled, vlined, linesnumbered]{algorithm2e}
\usepackage{bm}
\usepackage{lettrine}
\usepackage{tabu}
\usepackage{longtable}
\usepackage{multicol}
\usepackage{longtable}
\usepackage{url}
\usepackage{placeins}

\newlength\myheight
\newlength\mydepth
\settototalheight\myheight{Xygp}
\usepackage{graphicx}
\usepackage{subfigure}
\usepackage{bm}
\usepackage{lettrine}
\usepackage{xspace}
\usepackage{authblk}
\usepackage{pbox}

\begin{document}

\title{Why is My Secret Leaked? Discovering Vulnerabilities in Device-to-Device File Sharing}


\author{Andrei Bytes\qquad Jay Prakash\qquad Jianying Zhou\qquad Tony Q.S. Quek}

\authorrunning{Bytes et al.}
%
\institute{Singapore University of Technology and Design}




\maketitle
\begin{abstract}

The number of active users of Wi-Fi Direct Device-to-Device file sharing applications on Android has exceeded 1.8 billion.
Wi-Fi Direct, also known as Wi-Fi P2P, is commonly used for peer-to-peer, high-speed file transfer between mobile devices, as well as a close proximity connection mode for wireless cameras, network printers, TVs and other IoT and mobile devices.
For its end users, such type of direct file transfer does not incur cellular data charges. However, despite the popularity of such applications, we observe that the software vendors tend to prioritize the ease of user flow over the security in their implementations, which leads to serious security flaws.
We perform a comprehensive security analysis in the context of security and usability, and report our findings in the form of 17 Common Vulnerabilities and Exposures (CVE) which have been disclosed to the corresponding vendors.
To address the similar flaws at the early stage of the application design, we propose a joint consideration of security and usability for such applications and their protocols that can be visualized in form of a customised User Journey Map (UJM).

\end{abstract}






\section{Introduction}




Device-to-Device (D2D) communication facilitates a direct connection and single hop communication between compatible radio-frequency (RF) devices without the need for association with access points (APs) or cellular base stations (BSs).
Modern D2D communication leverages high data rate and is beneficial for mobile-to-mobile file sharing, wireless printing, screen-casting, and a wide range of other applications.
With the introduction of Wi-Fi Direct by Wi-Fi Alliance and its integration by Google into Android 4.0 \cite{android41direct}, its user base and use cases have increased exponentially over the past years \cite{D2D}. 
The number of active users of Wi-Fi Direct D2D file sharing applications on Android has exceeded 1.8 billion.

The use of Wi-Fi Direct provides significant usability benefits, as compared to Peer-to-Peer (P2P) communication over conventional Wi-Fi access points.
This is mostly due to two reasons: a) high speed over Wi-Fi direct D2D RF link, and b) straightforward, easy to use flow of pairing and data exchange. 
Since usability plays significant role in adoption of mobile applications by people, simple and easy to use interactions are preferred \cite{pig}. However, we will show how usability has been misused at cost of security in top Android application for file sharing.

In this paper, we performed a practical \textit{security-usability analysis} of the most downloaded D2D mobile sharing implementations, and reported the findings to the corresponding vendors.
We highlighted the causes of those vulnerabilities and suggested the usability-security trade-offs to avoid those vulnerabilities in protocol design.
We also quantified a \textit{combined notion of usability and security} which could help the protocol designers to evaluate the risks of usability-security trade-offs being adopted at the early stage of protocol design.


The rest of this paper is organized as follows. In Section 2, we provide short background on Wi-Fi Direct and D2D file sharing applications and the problem state. In Section 3, we present our methodology for security and usability analysis of the most downloaded D2D mobile sharing implementations. In Section 4, we group the identified vulnerabilities by common types and discuss their usability context. In Section 5, we propose a methodology of mapping the system design decisions into the User Experience space as an attempt to address similar insecure design decisions at the early stage. Related work is reviewed in Section 6 and Section 7 concludes the paper.

\vspace{-4mm}
	
	\section{Background}

	\subsection{Wi-Fi Direct}

	\begin{figure}
		\centering
		\includegraphics[width=.6\textwidth]{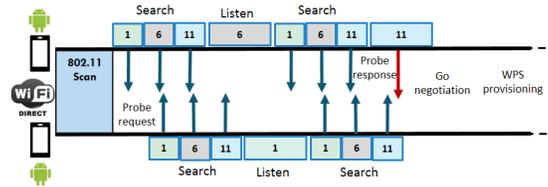}
		\caption{Typical discovery and WPS provisioning in Wi-Fi Direct \label{802}}
	\end{figure}

A typical Wi-Fi Direct link consists of a central device called as group owner aka GO, which has all functionality like an AP, and other connected device(s) are referred as client(s). A successful group formation occurs in three phases, \textit{Device Discovery, Service Discovery,} and \textit{WPS Provisioning} \cite{wps}. As shown in Fig.~\ref{802}, devices switch themselves between two states, the \textit{search state} and the \textit{listen state} in discovery phase. In the search state, the device sends a probe request on a channel, either of channel 1, 6 and 11, and waits on the same channel for the response for a fixed time, \textit{dwell time}.  Then it jumps to another channel and performs the same operation. After completing search operation on all channels, the device switches itself to listen state and remains on any one of the social channel, while listening for probe requests. It replies back with probe response after encountering any probe requests. A successful device discovery occurs when a probe request and the corresponding probe response exchange, with other device in proximity, takes place on the same social channel. Post first two phases, WPS provisioning facilitates a secure connection.

	\subsection{D2D file sharing applications on Android}

     \textbf{Google Files} \textit{(com.google.android.apps.nbu.files)} 
     is referred as ``Files by Google'' and ``Google Files Go''. Originally developed within the Google's Next Billion Users (NBU) \cite{nbu,nbu1}-the latest generation of internet users to come online on smartphones in places like Brazil, China, India, Indonesia and Nigeria- project to target emerging markets, it is being actively endorsed on Android and comes pre-installed as a system application since Android 8 ``Oreo'' and Android 9 ``Pie'', as well as Android Go editions for lower-end devices \cite{androidgo_description}. \textbf{SHAREit} \textit{(com.lenovo.anyshare.gps)} was launched by Lenovo in 2015 and has quickly become a world's most widely used D2D file sharing application. As claimed by the vendor, the current number of active users of the application on desktop and mobile devices exceeds 1.8 billion \cite{shareit2018about}.
	It was reported in \cite{sensortower2018report} that by the end of 2017, SHAREit reached \#5 worldwide ranking position by number of installations among non-game applications. Shortly after the early version was released, multiple vulnerabilities related to weak security policies and weak passwords were reported
	\cite{cve2016shareit}. From the static analysis we have observed that the latest version still contains significant parts of the vulnerable legacy codebase. Notably, despite some functionalities are no longer referenced in the UI, they can still be triggered remotely through the built-in embedded webserver.

	The Android versions of \textbf{Xender} \textit{(cn.xender)},  \textbf{SuperBeam}\\ \textit{(com.majedev.superbeam)}, \textbf{Zapya}~\textit{(com.dewmobile.kuaiya.play)}, and \textbf{MiDrop} \textit{(com.xiaomi.midrop)} are another commonly used device-to-device file sharing applications that actively compete with SHAREit, with more than 300 million installs on Google Play. Table \ref{table:google_play_stats} listed the D2D file sharing applications to be analysed in this paper. 


		\vspace{-5mm}
			\begin{table}[]
			\centering
		\begin{adjustbox}{width=0.55\textwidth}
		\begin{tabular}{|l|l|l|}
			\hline
			\multicolumn{1}{|c|}{\textbf{Application}} & \multicolumn{1}{c|}{\textbf{Package name}} & \multicolumn{1}{c|}{\textbf{\# 
					of installs}} \\ \hline
			SHAREit                              & com.lenovo.anyshare                        & \textgreater 1 Billion                            \\ \hline
			Xender                               & cn.xender                                  & \textgreater 100 Million                            \\ \hline
			Xiaomi Mi Drop                       & com.xiaomi.midrop                          & \textgreater 100 Million                            \\ \hline
			Files by Google                       & com.google.android.apps.nbu.files                          & \textgreater 100 Million                            \\ \hline
			Zapya                                & com.dewmobile.kuaiya.play                  & \textgreater 50 Million                             \\ \hline
			SuperBeam                            & com.majedev.superbeam                      & \textgreater 10 Million                             \\ \hline
		\end{tabular}
		\end{adjustbox}
		\caption{Shortlisted applications for our analysis}
		\label{table:google_play_stats}
	\end{table}\vspace{-17mm}

    \subsection{Practical limitations}
    A few significant technical challenges appear in practical implementations of D2D file sharing systems. 
    
    \textbf{Usability of Wi-Fi Direct:} Essentially, Wi-Fi Direct protocol does not have a dedicated way to agree on a shared secret. The Wi-Fi Protected Setup (WPS; originally, Wi-Fi Simple Config) was introduced to facilitate the secure association using either PIN or push-button confirmation. The usability goal was to enhance the flow as the majority of users are not comfortable with configuration dialogues and embedded devices which might not have peripherals to provide a setup interface. The WPS architecture also supports a Registrar - either a separate device or integrated to the AP service which helps client devices in enrolling to the network. The short PIN is commonly used to agree on secret keys \cite{d2d1}. While push-button is vulnerable against nearby attackers, short numeric PIN can be exposed and guessed in multiple ways \cite{bongard2014offline}, \cite{wps2014HackLu}. If the association flow allows the user to set their own PIN, this adds a human factor to the system, causing a risk of using predictable numeric combinations. 
    
    \textbf{Compatibility with legacy devices:} 
    Due to compatibility reasons in supporting connections with legacy devices, the common mode of D2D file sharing on Android is through the setup of a Wi-Fi P2P group \cite{android_Wi-Fi_p2p}. The first device serves as a Group Owner (GO), while one or more devices connect to the network as clients.
    In this mode, a traditional WPA2 passphrase is set by the Group Owner. The legacy clients can connect to the group even if they do not have Wi-Fi Direct support. 
    
    \textbf{Shared secret agreement:}
    Both Wi-Fi AP and Wi-Fi P2P group modes have a fundamental problem of the secure generation and transmission of network credentials (most commonly, the WPA2 pre-shared key) among the devices.
    This leads the problem of secure association.
    
    \textbf{Scanning in Wi-Fi:}
    The passive and active scanning capabilities of the Wi-Fi stack does not fit all use cases in mobile file sharing. Thus, from the user experience point of view, during a file sharing session, finding the right channel and association consumes more time than the actual transmission of a single document. Furthermore, the  channel probing is often initiated before the target network is up, causing noticeable delays in the user flow. 
    
    \textbf{User experience in association:}
    By design, the authentication mechanisms in the underlying protocols used in D2D file sharing applications require the user to produce certain input (entering the passphrase, pressing the WPS button) to achieve secure association. This complicates the creation of the seamless interface and user flow of the file sharing application.
    
    \textbf{Incoming confirmation:}
    Similarly, if the application offers to directly verify the PIN (e.g. Bluetooth Secure Simple Pairing) code or other unique identifier of the peer, the user tends to skip this step due to short-time, ad-hoc nature of mobile file exchange and additional cognitive load. This raises the challenge of performing the file transfer confirmation in a simple and secure manner.
    
    \textbf{Encryption at the application layer:}
    While some encryption is normally provided by the link layer, the additional challenge is to address in-network confidentiality and integrity attacks against transmitted data.
    This may introduce extra complexity in building Public Key Infrastructure (PKI) between devices, especially in the absence of SSL/TLS for HTTP connections as described in Section \ref{insecurity}.
    
    $\blacksquare$\textbf{Research questions:} Based on the discussed limitations this paper motivates 3 research questions towards usable and secure solutions for D2D mobile file sharing: \textbf{RQ1} How secure is the implementation of the most commonly used D2D file sharing applications on Android? \textbf{RQ2} What key usability factors are behind the insecure design decisions? and \textbf{RQ3} How to correlate vulnerabilities in usability space to address them at early design stage?
 


	\section{Analysis}
	
	In this section we analyze the six most downloaded Wi-Fi Direct mobile file sharing applications to demonstrate the correlation between usability and security of their implementations. Based on the download statistics from Google Play, we have selected most popular file sharing applications as listed in Table \ref{table:google_play_stats}. 
	\vspace{-2mm}
\subsection{Methodology and setup}	
	Wi-Fi Direct based file sharing applications are widely used and retain a large user base due to the simplicity of the D2D connection setup and high transfer speeds. However, during our analysis we have identified a number of workarounds and security violations that vendors introduced in their products to untangle the user experience and gain access to a wider market share. As we show later in this section, the applications from our shortlist commonly introduce default or predictable connection credentials for seamless association between peers or transmit the passphrases through side channels. At the same time, our analysis shows that the shortlisted applications prioritize the performance and compatibility over security at multiple layers of their implementation. \\

\textbf{End goals.}~We combine both static and dynamic vulnerability analysis techniques and automate the comparative execution analysis on multiple Android API platforms to achieve the following goals:
	
\begin{itemize}
  \item Determine which network protocols are used for D2D file sharing.
  
  \item Locate the corner execution paths which might drop encryption of the communication (e.g. switching to unprotected wireless AP instead of keeping the Wi-Fi Direct link).

  \item Locate exploitable flaws which enable attacks by Receiver against Sender and vice versa (e.g. command/content injections, vulnerabilities in built-in web servlets).
  
  \item Locate remotely exploitable flaws which allow for private data leakage by third-party attacker through device network interfaces.
  \end{itemize}
  
\textbf{Reverse engineering.}~Statically, for more detailed manual analysis we fingerprint the execution paths which contain signs of the identifiable patterns.
  
	\begin{itemize}
  \item References to known Java and native components and libraries (e.g. NanoHTTPD in Xender).
  
  \item Calls to sensitive or unusual Android APIs (remote storage binding, process monitoring, command execution).

  \item Hard-coded values of certain format (URI schema fragments, regular expression templates, long number sequences, tokens, hashes).
  
  \item Potential misconfiguration of Android-specific components (permissions, exported Activities, exposed binding of Intents, Services triggered with poor validation).
  
\item Cryptography-related operations, random string generation, string encoding methods.

\item Implementation of remote and local URI validation (schema matching, regular expressions).

\item Conditions which tend to change the execution flow to fit a particular Android OS version or device vendor-specific APIs.
  \end{itemize}
  
 \textbf{Dynamic analysis.}~This is further extended with dynamic analysis to inspect the insecure behavior with the following methods: 
  \begin{itemize}
  \item Mapping of embedded endpoints to actual D2D sharing code snippets.
  \item Tracing the uses of network sockets when certain functionality is requested.
  \item Hooking Java and Android API interfaces to inspect call arguments, specifically where the code relates to cryptography operations, wireless AP configuration, Bluetooth discovery routine.
  \item Generating a word-list which includes the names of sent, marked for sharing and received files, IPv4 and hardware addresses involved in communication to filter out the method calls in execution flow.
\end{itemize}

  Application of these procedures in comparative execution analysis of identical scenarios on multiple physical and emulated devices allowed to identify certain discrepancies in the functional behaviours for deeper manual investigation of each case.
  
  The corner behaviour cases normally occur when certain functionality is not supported by the device or is not permitted in the current Android API. Notable effects of the latter include fallbacks to unprotected wireless APs from Wi-Fi Direct, switching to hard-coded credentials and setup of insecure limitations for credentials length in particular execution environments.

	\vspace{-2mm}
    \subsection{Typical implementation of discovery and pairing}\label{modes}
    A common example of the device discovery and pairing flow is the scheme used in Files by Google application on Android.
    The process can be divided into the following stages which are required to achieve successful pairing and connection verification:
    
     \begin{itemize}
     \item Sender generates a numeric connection ID.
     
     \item Receiver starts broadcasting a specially encoded hostname through Bluetooth API.
     
     \item Sender scans the nearby Bluetooth devices, finds and decodes back the Receiver's username from her hostname.
     
     \item The parties perform Bluetooth Secure Simple Pairing with a standard 6-digit verification code.
     
     \item The 6-digit verification code is blindly accepted by the Receiver device without user action.
     
     \item A confirmation dialog is displayed on the Receiver side.
     
     \item After user confirmation, the Receiver device raises either Wi-Fi Direct Group as an owner or Wi-Fi AP.
     
     \item The PSK is transmitted through the active Bluetooth link.
     
     \item Sender de-associates the Bluetooth adapter and connects to the WPA2 network.
     \end{itemize}
    
    
    
    Protocols implemented by vendors for Device-to-Device pairing vary depending on the mode of operation, user settings and compatibility with the device firmware. The comparison of scenarios, identified in the shortlisted application can be found in Table \ref{table:pairing}(in Appendix). 

	 

\vspace{-4mm}
	\subsection{Encryption and network protocols}\label{protocols}
 Despite wide advertisement of encrypted transmission channel in the product descriptions, it is observed that the analyzed applications rely on the link layer for confidentiality of the transmitted data. 
    Table \ref{table:transport} lists the use of known protocols by the applications and the exposed network ports.
	\vspace{-6mm}
		\begin{table}[]
		\centering
		\begin{adjustbox}{width=0.55\textwidth}
			\begin{tabular}{|c|c|c|c|c|}
				\hline
				\textbf{Application} & \textbf{Version} & \textbf{Protocol used} & \textbf{Ports used} & \textbf{Encrypted} \\ \hline
				SHAREit & 4.5.84 & UDT\footnote{http://udt.sourceforge.net/} & 52999 (UDP) & No  \\ \hline
				Xender & 5.1.1.Prime & HTTP & 6789 & No   \\ \hline
				Xiaomi MiDrop & 1.22.4 &  TCP; FTP & Random; 2121 & No \\ \hline
				Files by Google & 1.0.220185905 & TCP & Random; 10061 & Yes  \\ \hline
				Zapya & 5.7 (US) & HTTP & 9876 & No   \\ \hline
				SuperBeam & 4.1.3 & HTTP & 8080 & No   \\ \hline
			\end{tabular}
		\end{adjustbox}
		\caption{Use of network protocols} \vspace{-5mm}
		\label{table:transport}
	\end{table}\vspace{-10mm}

\vspace{-2mm}
	\subsection{Embedded HTTP endpoints and alternative sharing modes}\label{web}
	In addition to the primary mode of file exchange, all applications in our shortlist except Files by Google contain an embedded web-server able to serve dynamic pages. The reason for this is the compatibility requirement for file sharing with desktop computers and low-end mobile devices. In the recent versions of SHAREit and Xender, we also notice a dedicated web applications optimized for low-end KaiOS \cite{kaios_architecture} devices. The web applications and their asynchronous APIs are reachable remotely through the network interfaces of the Android device. In our analysis, we paid close attention to review these endpoints. The first goal is to identify the vulnerabilities, which are typical to desktop web applications and in this way are also brought into mobile space. Secondly, the review of the codebase shows that each of this applications, which can be identified by the unique TCP port it is served on (Table \ref{table:transport}, implements its own access control mechanism, while sharing the common resources and database with other endpoints. Such specifics of the architecture enabled us to identify attack scenarios where multiple vulnerabilities in separate embedded web applications, triggered sequentially through separate TCP ports can be chained to bypass the access control logic. The result of such architecture flaw causes leak of user files, unauthorized actions and remote upload of malicious files to the victim`s mobile device (More at Section \ref{insecurity}).

	
	
	
	
	\vspace{-4mm}
	\section{Vulnerabilities}\label{insecurity}
	In this section, we summarize our findings, which are common for the analyzed applications and discuss the usability context behind them. Notably, during the analysis of shortlisted targets we have observed that vendors tend to mirror the user flow and implementation patterns of each other. Partly, the reason for this is the nature of competition for the large existing user base, which resides on the same platform. A radical change in user interface or implementation of additional security features can create competitive disadvantage and thus is generally avoided. We note that the reflection of identical user interface and interaction flow (pairing, transfer confirmation) tends to spread security vulnerabilities which appear to be common for multiple vendors.
	\vspace{-8mm}
		\begin{figure}
		\centering
		\subfigure[SHAREit for PC: Guidance to use default password]{\includegraphics[scale=0.47]{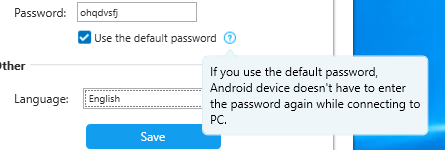}\label{fig:shareit_pass_settings}}\hfill
		\subfigure[Xender: Setting unprotected hotspot ]{\includegraphics[scale=0.45]{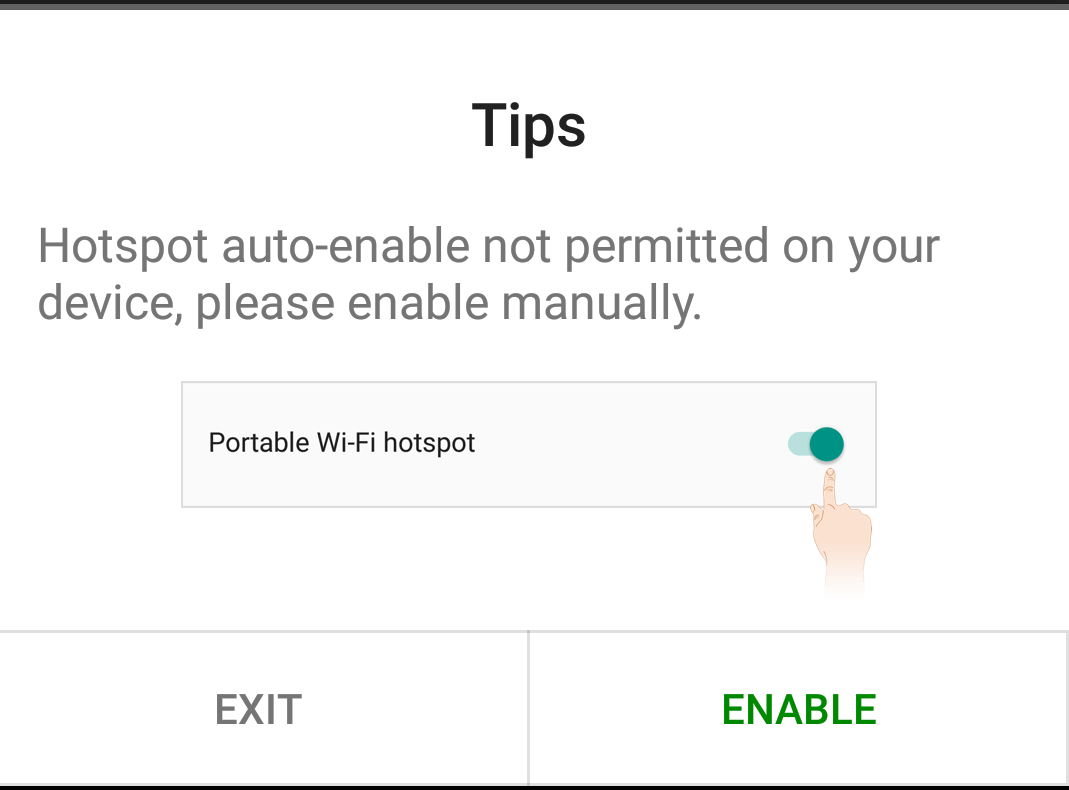}\label{fig:71_xender_1}}\vspace{-4mm}
		\caption{Insecure user flow in SHAREit and Xender}
		\vspace{-2mm}
	\end{figure}

\vspace{-10mm}
	\begin{figure}
		\centering
		\subfigure[SHAREit: Switching to Wi-Fi AP]{\includegraphics[scale=0.34]{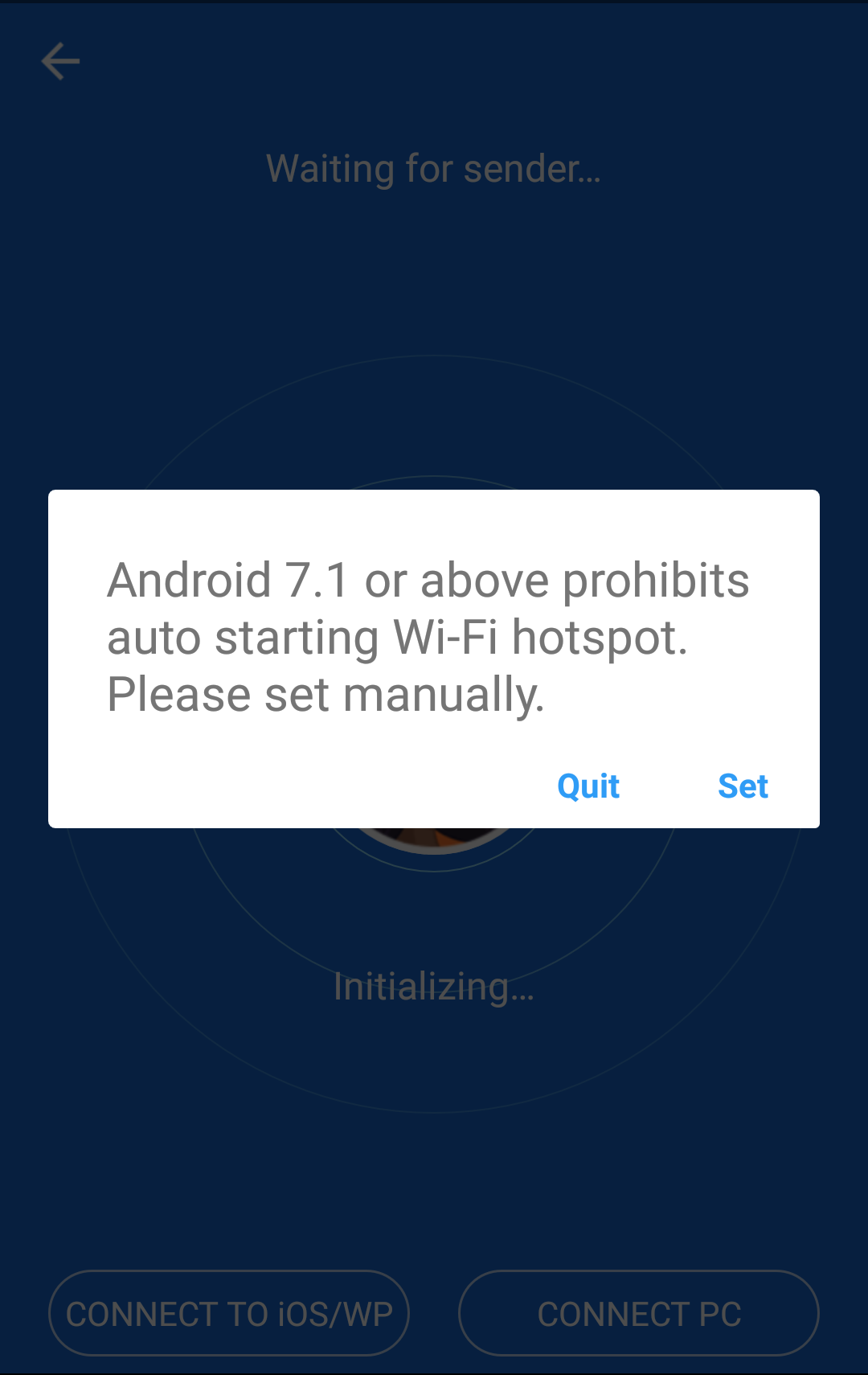}\label{fig:71_shareit_1}} \hfill
		\subfigure[SHAREit: Hotspot security mode is reset to None]{\includegraphics[scale=0.34]{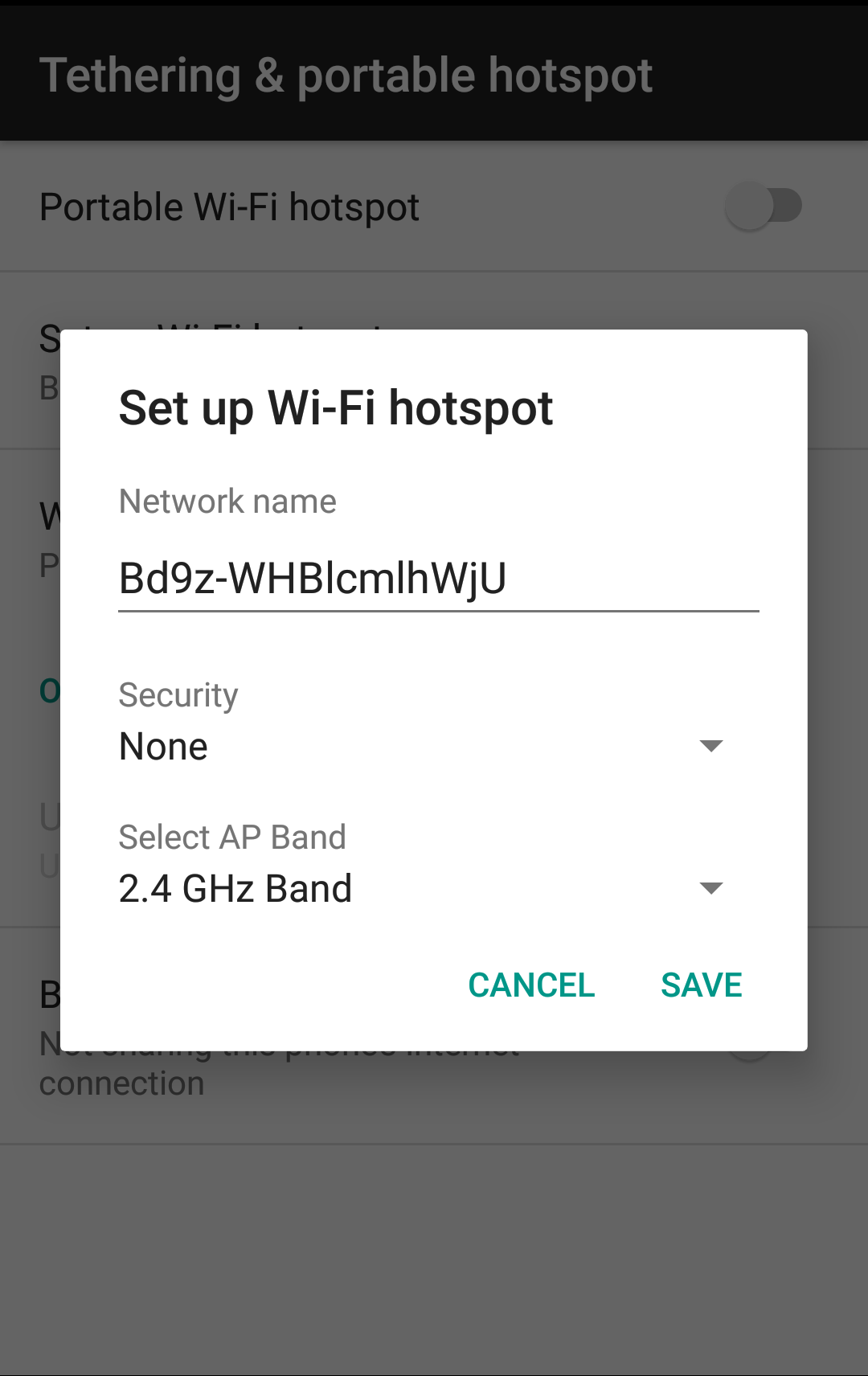}\label{fig:71_shareit_2}}\hfill
		\subfigure[Zapya: Preventing the passphrase setting for the hotspot]{\includegraphics[scale=0.34]{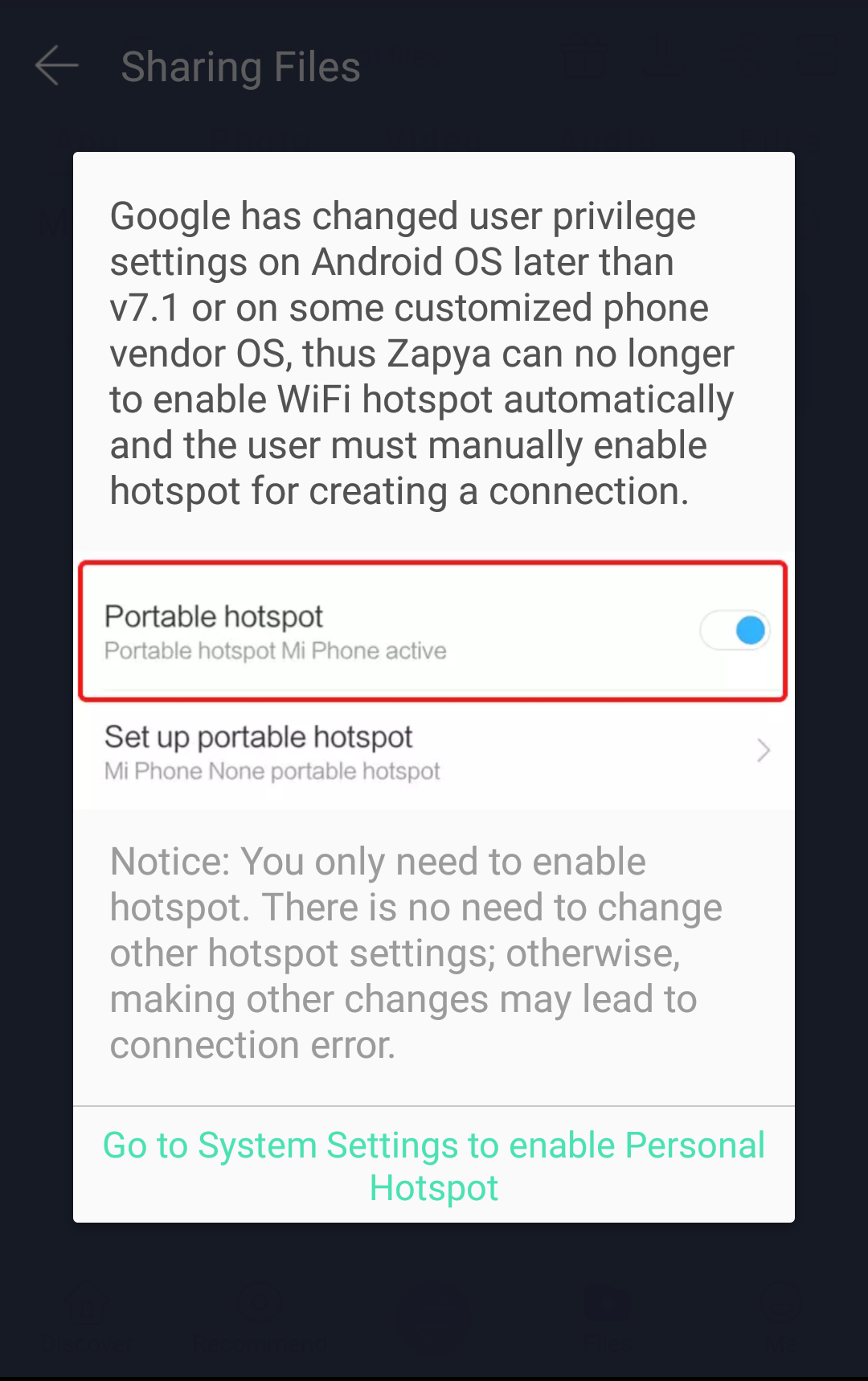}\label{fig:71_zapya_1}}\hfill
		\vspace{-4mm}
		\caption{Common usability favors}\vspace{-4mm}
	\end{figure}
	
	A key property, by which we have picked applications for our analysis was their advertised use of Wi-Fi Direct for D2D file sharing.
	Surprisingly, it was observed that every application in our shortlist, including Google Files 1.0.220185905 for Android implements additional fallbacks and is not using Wi-Fi Direct at all times. 
	In particular, a common behaviour for the analyzed applications is to silently turn either the Sender or the Receiver into a conventional Wi-Fi Access Point, often disabling the authentication or sharing a hard-coded default passphrase.
	The user, in her turn, is not informed of such behavior in most cases and expects the files to be sent through an encrypted Wi-FI Direct link.  We highlight the additional security impact, introduced by these fallbacks in our findings.

	%
	%

$\blacksquare$ \textbf{Usability of Authentication:}
	A number of key design decisions has been made in the reviewed applications with a clear priority put on seamless device discovery and association of peers. To achieve the effortless interaction flow, the application vendors commonly build custom association schemes. If either Wi-Fi AP or Wi-Fi Direct Group is established, there is no prescribed way to securely generate the secret and share it with the connecting clients. Thus, we observe a wide use of default credentials or passphrase generation algorithms which are built in a way that they can be independently derived by the client. In addition to these flaws, we have also noticed that the authentication behavior of the identical application version can vary when executed on different Android environments. The unexpected effects we have reported include switching to side channels to send a passphrase and the AP name and forcing the user to set completely unprotected AP manually, in the Android settings. Similarly, in order to relieve the user from the obligation to enter the password, the applications tend to set insecure pre-defined settings without clear notification shown to the user. A notable example of such is the the association flow of SHAREit on Android with its desktop companion application of version 4.0, which allows file exchange with Windows hosts. Upon start-up, the desktop application immediately raises the AP, using a hard-coded passphrase. The notification of this is not shown to the user. The user has no control to read or change the passphrase at this point. Next, the client Android device independently predicts the credentials in order to associate with the desktop application automatically with no pre-shared secret but using a hard-coded derivation algorithm. At the same time, the UI of the desktop application offers a rather obscure location of the settings dialog, hiding it behind a click on the user avatar. Even though there is a field in the application settings to change the AP password for next session~\ref{fig:shareit_pass_settings}, a warning message discourages a user from changing it. The user is advised to use the default WPA2 passphrase in exchange for keeping the connection process simple. Unfortunately, no explanation on the security consequences of such setting is provided.
	
	Alike SHAREit, other applications from our shortlist were identified to use similar insecure workarounds to simplify the user flow and the authentication of the Sender and the Receiver. The descriptions of these issues are listed in Table \ref{table:cve_list}.

	$\blacksquare$ \textbf{Performance over Security trade-offs}
	Keeping in mind the wide presence of authentication trade-offs in the reviewed applications which can facilitate the attacker in gaining access to the network, we have consequently examined the implementation of data transmission in the established network between the Sender and the Receiver. As was previously mentioned, even though all applications in our list declare Wi-Fi Direct as their primary way of association, in practice this is not always the case. Due to the automatic selection of fallbacks which is normally out of user control, the data transmission happens through an unprotected Wi-Fi AP or within the existing Wi-Fi connection. 
	Thus, the functionality which is designed to rely on the encryption, provided by the network layer by Wi-Fi Direct is instead exposing the user transmitted files in clear-text though non-encrypted transport protocols (Table \ref{table:transport}). SHAREit uses UDT~\cite{udtprotocol} protocol and relies solely on the network layer security configuration, thus lacking any additional encryption or integrity protection for transferred files.
	
	Another notable case which commonly results in user data being transmitted over existing network connection or the AP in clear-text is the auxiliary functionality of the shortlisted applications. Thus, Xiaomi MiDrop introduces a "Connect to PC" feature which is different from its primary mode of operation. Our analysis showed that in this mode the application exposes unrestricted access to the device filesystem by acting as an FTP server. The server does not isolate the file exchange folder nor uses any authentication by default. The FTP connection is served to an anonymous in-network user. Naturally, this solution has no encryption at the application layer and the port is exposed in any network that Android device is associated to, regardless of the type of this underlying link.

	Except for Google Files, all the reviewed applications also support a Web Sharing mode which allows to exchange files over HTTP with other peers. It was observed that in this mode none of the applications which we have reviewed provide SSL \textbackslash{} TLS or any other option to protect the confidentiality of the transferred files, exposing the communication to an in-network attacker in clear-text (Table \ref{table:transport}).
	Additionally, the embedded web server functionality introduces additional security vulnerabilities, delivered by its custom endpoints. We further these in the next paragraph.
	
 \begin{table}[]
	\centering
	\begin{adjustbox}{width=1.1\textwidth}
		\begin{tabular}{|m{0.2\textwidth}|m{0.3\textwidth}|m{0.3\textwidth}|m{0.3\textwidth}|}
			\hline
			\textbf{Application} & \textbf{UI feature name} & \textbf{Discovery} & \textbf{Pairing} \\ \hline

			Files by Google & Share - Send & Programmatic (Bluetooth scan) &
			WPA2 PSK over Bluetooth. Custom 6-digit connection ID confirmed by the Receiver  \\ \hline
			
			SHAREit & Send & Programmatic (BT scan) & QR code at Receiver (12 byte PSK)  \\ \hline
			
			SHAREit & Send - Connect to iOS & Manual (Wi-Fi AP) & Type in 12 byte PSK \\ \hline
			SHAREit & Send - Connect PC & Manual (Shared network, Web URL)  & QR at client (desktop application) side  \\ \hline
			SHAREit & Share with  KaiOS & Manual (Wi-Fi AP) & Hard-coded PSK derivation logic in KaiOS client \\ \hline
			SHAREit & Connect PC & Manual (QR and Web URL) & QR at client (desktop application) side  \\ \hline

			Xender & Send & Programmatic (BT scan) & QR at Receiver (12 byte PSK)  \\ \hline
			
			Xender & Connect PC & Manual (Shared network, Web URL) & Confirmation dialog at Receiver    \\ \hline
			Xender & Connect KaiOS & Manual (Wi-Fi AP) & Derived PSK derived or Type in 12 byte PSK \\ \hline
			Xender & Scan Connect & Manual (QR) & QR at Sender  \\ \hline

			Xiaomi MiDrop & Send &   \shortstack{Manual (QR)}  & QR and Confirmation by the Receiver (6-digit ID)  \\ \hline
			Xiaomi MiDrop & Connect to Computer & Manual (Shared network, FTP hostname) & Unprotected (Public FTP share)  \\ \hline
			Xiaomi MiDrop & Webshare &  Manual (WiFi AP and Web URL) and Web URL & Type-in 12 byte PSK \\ \hline

			Zapya & Send & Manual (QR at Receiver) & QR or type in passphrase \\ \hline
			Zapya & Group Share &  Manual (QR at Sender) & QR or type in passphrase \\ \hline
			Zapya & Send - Bluetooth Assist &  Programmatic (BT scan) & WPA2 PSK over BT  \\ \hline
			Zapya & Shake to Connect & Mixed (BT scan initiated by hardware sensor event) & WPA2 PSK over BT \\ \hline
			
			Superbeam & Send - Legacy & Programmatic (NFC) or Manual (QR) & QR or type in 118 byte key  \\ \hline
			Superbeam & Send - Secure &  Programmatic (NFC) or Manual (QR) & QR or type in 32 byte key  \\ \hline

		\end{tabular}
	\end{adjustbox}
	\caption{Discovery and pairing modes}
	\label{table:pairing}
\end{table}

$\blacksquare$ \textbf{Legacy code and vulnerable servlets:}
	SHAREit, at its early versions, has been actively engaging a built-in web server functionality. At its current version (4.5.84), this functionality is still present in the application but is mainly used as a fallback to communicate with desktops and mobile devices running platforms different from the host. Our static analysis of the SHAREit 4.5.84 for Android showed that a major amount of legacy functionality is not used in the user interface anymore yet is still served by the web server, exposing a number of endpoints, that can be remotely triggered from any network that the device is associated to. The implementation of this code, including the code-base, currently used to support the Web share feature, has poor access control mechanisms and often lacks input sanitation, allowing the attacker to ex-filtrate files from the device and perform Cross-Site-Scripting (XSS) against the Receiver.
	Thus, it was observed that multiple endpoints of Superbeam Web share mode does not sanitize the input data, allowing for injecting reflected and stored XSS, performed by the Sender. For the latter, a stored payload is rendered into the UI from the filenames, which the Sender advertises and is rendered from them on the Receiver`s side.
	
\begin{figure}
	    \centering
	    \includegraphics[scale=0.8]{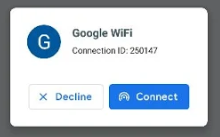}\vspace{-4mm}
	    \caption{Google Files: Manipulated association dialog}
	    \label{fig:goog_receiver_dialog}
	    \vspace{-6mm}
	\end{figure}

	Another scenario of a Sender-to-Receiver attack was observed in the Google Files application on Android. Due to the lack of parameter filtering and sanitation on both sides, it was possible for the Sender to transmit her crafted username over the network, which allowed to manipulate the contents of association confirmation dialog at the Receiver side, by rendering additional layout elements and commenting out the unwanted fields. An example of a crafted file transfer confirmation dialog, with the removed supporting text through the injection of an open comment tag, is shown in
	Fig.~\ref{fig:goog_receiver_dialog}.
	\vspace{-6mm}
	Similar to SHAREit, Xender application also provides a Web share feature for compatibility with desktops and third-party mobile devices. However, as opposed to SHAREit which only activates its web-server on port 2999 during file sharing, Xender immediately starts it in the background with the application runtime on TCP port 6789 of Android device, even when no file sharing is in process. Our study of the reconstructed application code and dynamic analysis identified an exposed endpoint, which allows to access arbitrary files from the device through file path manipulation.
	Regardless of the user`s intention to send or receive files, the vulnerable service is raised automatically and is exposed to anyone in the same network. This provides a stealth channel to obtain arbitrary files from the file-system without user notification, acting as a remote backdoor on the victim Android device.
	
	$\blacksquare$ \textbf{Password transmission through side-channels:}
	While MiDrop and Google Files rely on Bluetooth for proximity search of their peers, other applications use it as a side channel to transmit the association credentials between the Sender and Receiver. SHAREit requires a granted access to Bluetooth "to increase user connection speed" (Fig. \ref{fig:BT_shareit_1}). However, we have observed that if the Bluetooth connection is successfully established, SHAREit uses it to transmit the Access Point credentials in its fallback mode and seamlessly associates with the peer device. Otherwise, if the credentials cannot be transmitted with Bluetooth, the Sender will be asked to authenticate with a passphrase. Notably, the fact of establishing a Bluetooth connection is not reported to the user and requires no pairing or other confirmation.
	On the contrary, Zapya makes the user aware about the transmission of AP credentials over Bluetooth and provides an implicit switch to disable this feature (Fig.~\ref{fig:zapya_settings}). Xender and Superbeam, in turn, engage QR codes as a primary way to exchange the credentials, needed for sender and receiver to associate. The example of such QR code is shown at (Fig.~\ref{fig:qr_xender}) and encapsulates the credentials in the URI, the AP name and its passphrase are observed at \textit{nm} and \textit{pw} parameters:
	\begin{verbatim}
	http://www.xender.com?nm=AndroidShare_4615
	&pw=049a0ae278e5&i=43&p=19638464
	\end{verbatim}\vspace{-3mm}
	\begin{figure}
		\centering
		\subfigure[SHAREit: "Bluetooth is used to increase user connection speed"]{\includegraphics[scale=0.34]{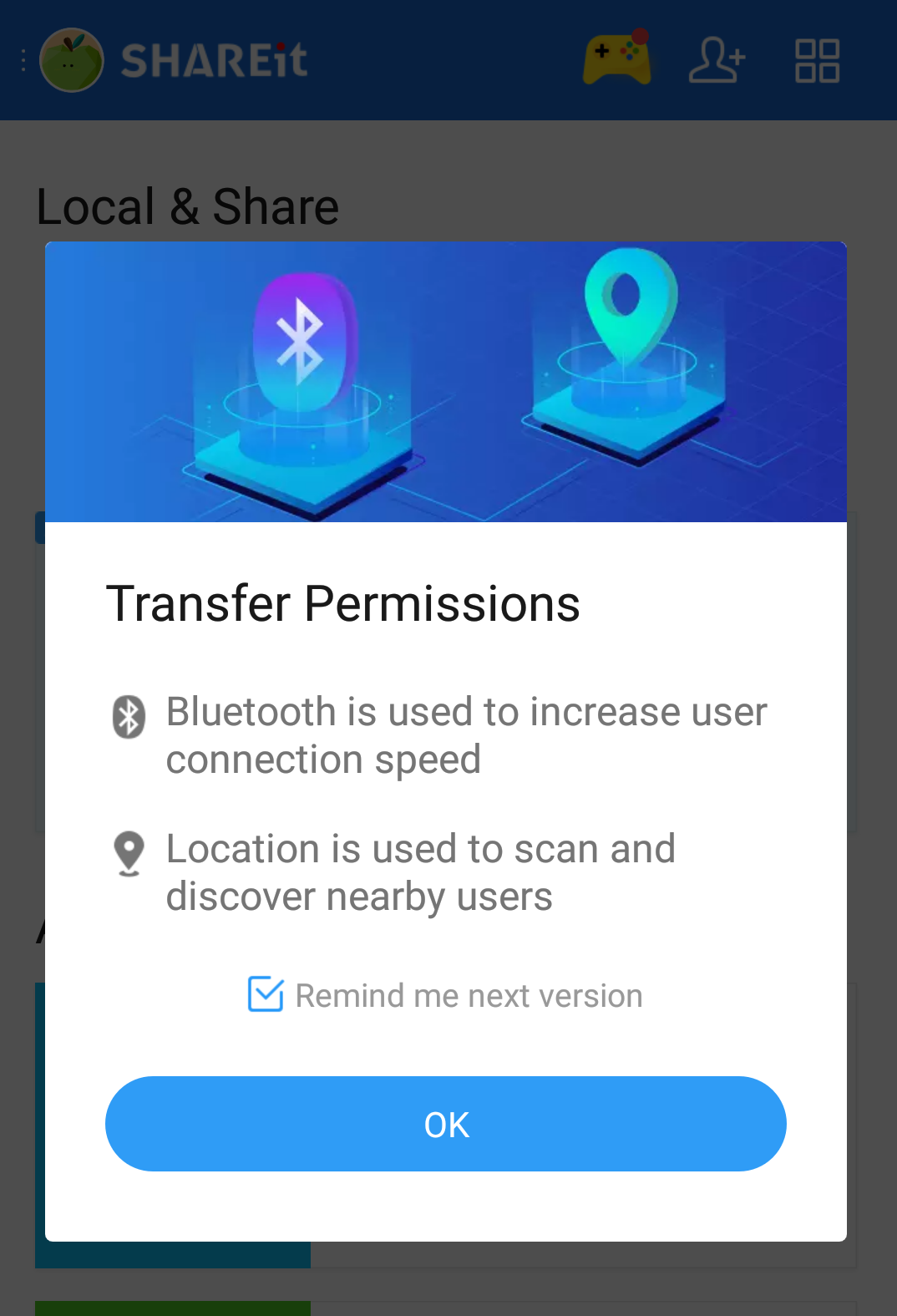}\label{fig:BT_shareit_1}}\hfill
		\subfigure[Zapya: Switch to disable credentials transmission over Bluetooth]{\includegraphics[scale=0.39]{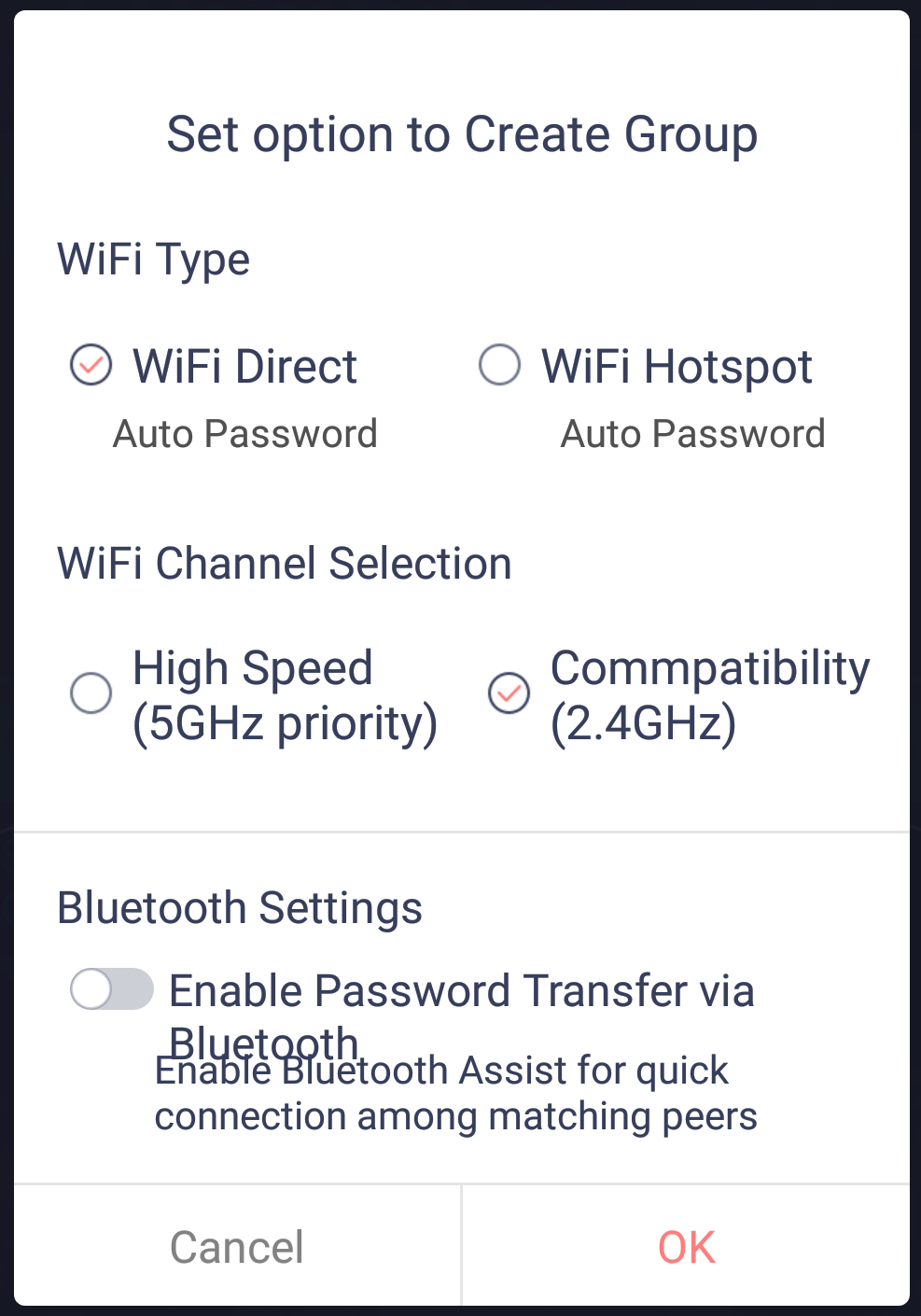}\label{fig:zapya_settings}}\hfill
		\subfigure[QR-encoded credentials in Xender]{\includegraphics[scale=0.32]{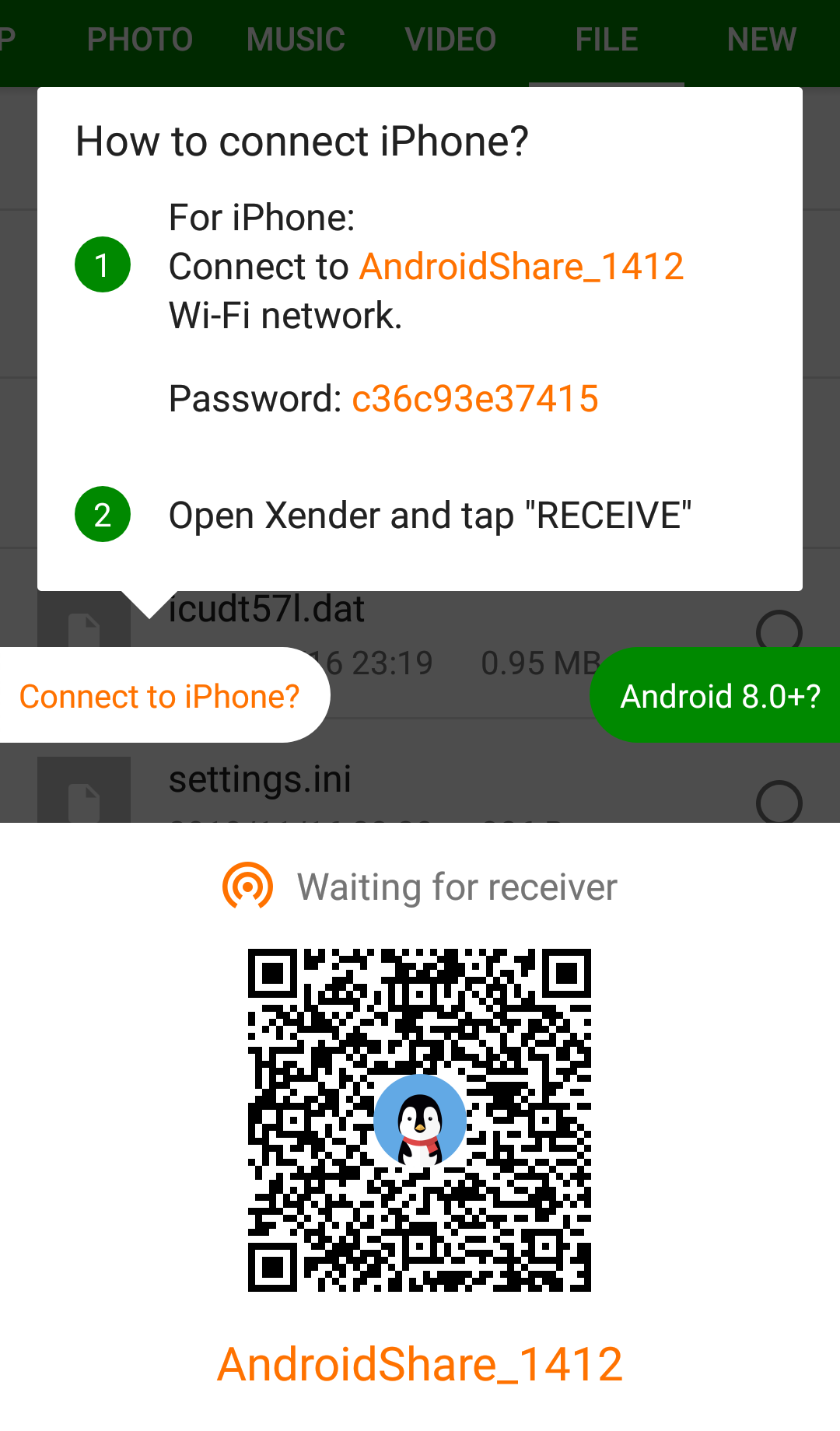}\label{fig:qr_xender}} \vspace{-4mm}
		\caption{Side-channel transmission of credentials}
		\vspace{-6mm}
	\end{figure}
	$\blacksquare$ \textbf{Insecure OS version-specific workarounds on Android 7.1 to 8:} The continuous deprecation of APIs in the Android security lifecycle \cite{mayrhofer2019android} often introduces additional permission restrictions for its non-system applications. With a natural intention to obtain more control on the application behavior and to improve general security and privacy posture of the platform, these can cause an unexpected effect for the users, causing the developers to urgently deploy workarounds.
	Thus, with the upgrade Android OS to 7.1, non-system applications lost the ability to programmatically raise a DHCP-enabled Wi-Fi Hotspot \cite{android71ticket} \cite{android71ticket2}. If the application is executed on newer Android APIs, Android 8 and 9, it can use an interface  \textit{Wi-FiManager.LocalOnlyHotspotReservation} which was introduced to particularly solve this problem \cite{android8hotspot}.
	
	Although on some devices and platform versions particular applications from our shortlist are shipped pre-installed with system privileges (Google Files, Xiaomi MiDrop), they do not always have this advantage. We have identified a common insecure workaround, specific to Android 7.1, implemented by most applications from our list. The efforts of developers to keep their applications functioning on this platform has resulted in solutions that override existing in-app security mechanisms which would be present if the application was executed with particular Android APIs.
	
	Thus, to keep the file transfer functioning when running on Android 7.1, ShareIT 4.5.84 , Xender 4.2.2.Prime and Zapya 5.7 (US) set the Android settings dialog with open AP (security: none) and ask for the users action to enable it (Fig.~\ref{fig:71_shareit_1}, \ref{fig:71_xender_1}). Moreover, in a case when the user pre-configures a hotspot with own WPA2 passphrase in Android settings, the above-mentioned applications would override these settings and permanently reset the security mode back to None (Fig.~\ref{fig:71_shareit_2}).
	Remarkably, Zapya even adds an explicit warning for the user to prevent her from making changes in the AP configuration: \textit{"Notice: You only need to enable hotspot. There is no need to change other hotspot settings, otherwise, making other changes may lead to connection error" } (Fig.~\ref{fig:71_zapya_1}). Indeed, ignoring this warning and manually protecting the hotspot with a password in the settings dialog resulted in complete malfunction of ShareIT 4.5.84 and Zapya 5.7. If the Access Point has WPA2 enabled, the peer is unable to authenticate and connect. Similarly, in Xender 4.2.2. Prime the connection dialog doesn`t allow to associate with its peer if its password is longer than 8 symbols. This limitation puts significant security limitations even when the user is concerned to encrypt her hotspot. Xiaomi Mi Drop applies an identical workaround for Android 7.1. However, instead of raising an unprotected host-spot, it sets a predefined password, which is programmatically predictable by the client. Changing this default password results in association failure, analogous to behaviour of SHAREit and Zapya.
	
	$\blacksquare$ \textbf{Deprecation of Wi-FiConfiguration in Android 10:}
	Notably, the initialization class for Wi-Fi networks faces another change in Android 10 (API 29) \cite{android10_wifi}. The creation of android.net.Wi-Fi.Wi-FiConfiguration which previosly was serving to set AP security mode and PSK is being replaced with Wi-FiNetworkSpecifier.Builder. Potentially, this can cause vendors to add even more routines to ensure the device pairing works programmatically on Android 10 or higher.

$\blacksquare$ \textbf{Reported vulnerabilities:}
	Table~\ref{table:cve_list} lists descriptions of vulnerabilities and assigned CVE IDs which we have reported to the corresponding product vendors, based on our findings, summarized above in this section.	\vspace{-3mm}

	\section{Correlation with UX space} \label{obs_state} 
\begin{figure}
 		\centering
 		\includegraphics[width=0.475\textwidth]{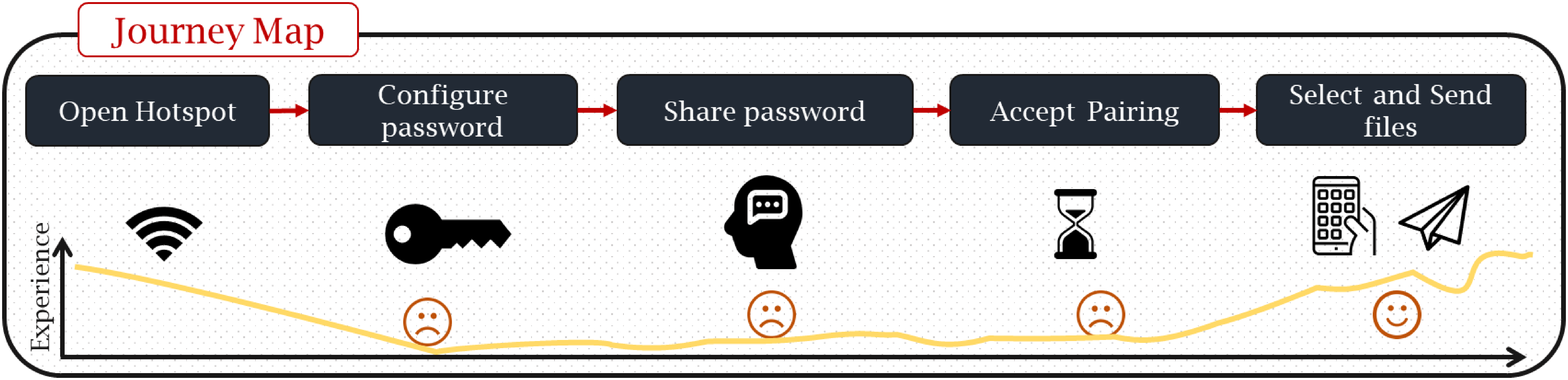}
 		\includegraphics[width=0.475\textwidth]{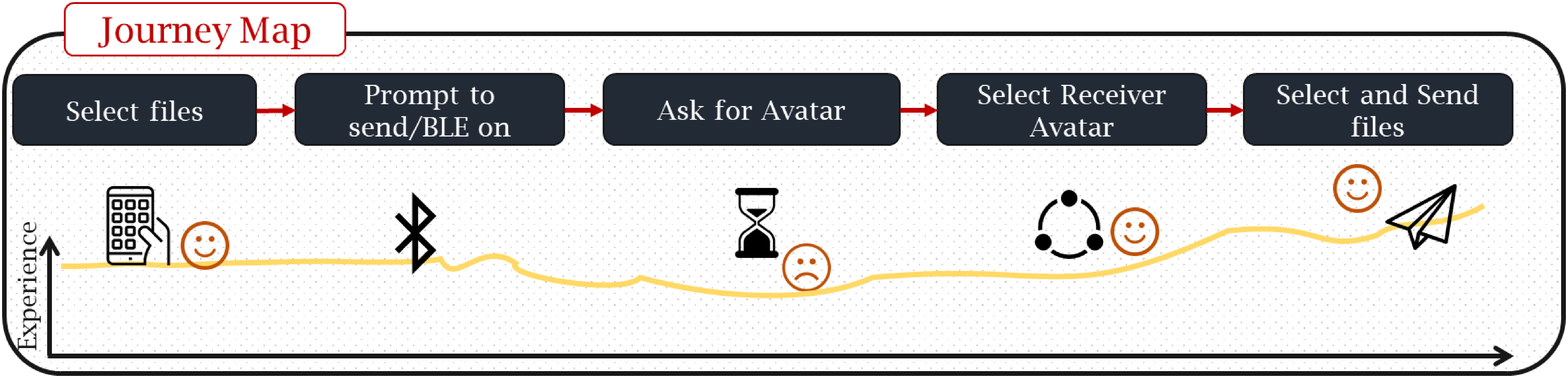}
 		\vspace{-4mm}
		\caption{User journey map for a) hotspot and b) ShareIt based file sharing \label{ujm_send}}\vspace{-8mm}
 	\end{figure}

\vspace{-2mm}
	In order to dig further in to the motives behind the vulnerabilities, we did usability studies with two groups: students (of engineering, design and research) and potential NBU users. Pairs were formed, among group themselves, and one of the member was asked to a share a set of 5 photos and 3 videos with the other. They had to repeat the transfer for all the selected file sharing applications and a typical hotspot method (where users are required to set in AP using a PIN and share the same to the receiver). Hotspot was taken as the benchmark for least usable solution. For understanding the overall experience and usability, we used system usability testing (SUS) \cite{sus} and user journey mapping (UJM) \cite{design_science},\cite{userexperience} and \cite{howard2014journey}. SUS is a subjective study and hence represents people's perspective of their experience which might be biased at occasions. Also, SUS returns a cumulative score, 100 being for an ideal solution, without insights on where actually the system lacks. Hence, SUS cannot serve purpose of usability diagnostic tool, \cite{sus_review}, which we are after. We compliment SUS with UJM because it helps in identifying pain points at each step and gives a representative picture of experiences as the steps are performed by a user. While users were performing the tasks, sequences of steps for file sharing, we noted their journey and also asked questions for mapping their user experience along the time series.  UJMs for hotspot and ShareIt are shown in Fig. \ref{ujm_send}(a) respectively. As can be seen, hotpsot based method has more pain points compared to ShareIt and so was the case with other D2D file sharing applications. These in themselves explain the reason behind popularity.
	\vspace{-6mm}
	
 		

			\begin{figure}
 		\centering
 	
 		\includegraphics[width=0.6\textwidth]{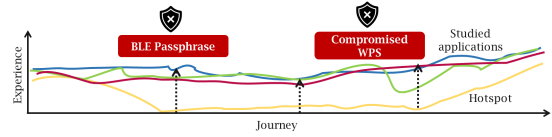}\vspace{-4mm}
		\caption{Correlation between usability and introduction of vulnerability in user's journey\label{correlation}} \vspace{-4mm}
 	\end{figure}\vspace{-8mm}


\subsection{What went wrong?}
For further diagnosis, we started identifying pain relievers in the file sharing applications, in terms of user experience and evaluate how they lead to introduction of vulnerabilities. The root causes have been discussed henceforth:

$\blacksquare$ \textbf{Usability Security Trade-off:} Fig. \ref{correlation}, plots user experience of popular file sharing applications as compared against hotspot-only (light yellow) based file sharing experience. As is obvious, a clear trade-off between security and usability comes in to play as economic gains drive UI and UX priorities. The experiences have improved but at the cost of vulnerable measures. The cause of such implementations are based in the fact that the initial authentication and secret establishment (generation and sharing of PINs) are not inherent part of the Wi-Fi Direct protocol. It is solely decided by the developers who try to re-invent the software flow. Developers, both UI and security, tend to work out a common solution based on a compromise between ideal solution based on HCI and security principles, \cite{sus1}.


$\blacksquare$ \textbf{Weak communication link between security designs and application developers:} For studying the links in usable security, we break the design of an end-to-end mobile application in to 3-phases connected with feedback loops, Fig.(\ref{changed}). Our investigation on application design process, in congruence with \cite{usable_security_book}, identifies two weak feedback links: a) feedback from the developers  and b) feedback from UX experts to protocol designers. The SUS and UJM based feedback to UI and application development team are well understood and taken care of in successive iterations of an application. But there exists little or no provision to convey issues to the protocol development team; due to different terminologies or parameters and weak communication channel, \cite{best_practice} and \cite{johny}. The paper rather proposes, whose basic outline is given in next subsection, to \textit{empower} security and protocol designers with an abstract understanding of takeaways by research communities, including findings from psychology, human-computer interaction, and design science.



$\blacksquare$ \textbf{Prevalence of usability studies primarily after complete design of applications:} UX expert review applications and products using different surveys and User Journey Maps (UJM) but only after deployments of software and user-interfaces (UI). 
The most frequently used tools like SUS caters primarily to usability and that too only after smartphone/desktop/web applications are ready with tentative UIs. Developers tend to use SUS metrics as prima facie of adoption and tend to ignore potential vulnerabilities. Some of the prior work on security and usability analysis of file-sharing applications, \cite{good2003usability}, suggest about vulnerabilities and set of guidelines for secure and usable design. \cite{usable_security_book} reviews large set of literature in usable security and privacy and \cite{design_with_usable_security} studies security and usability as two antagonistic goals and present design principles and patterns to achieve a jointly optimised goal. But their arguments are primarily focused on user control and privacy. A comprehensive and  dedicated framework of secure data sharing is needed.

\vspace{-4mm}	
\subsection{Addressing RQ3: Discussions to fix usability-security trade-off:} \vspace{-2mm}
$\blacksquare$ \textbf{Joint notion of usability and security:} Based on our findings we would like to encourage an interactive unification of system protocol and user spaces which can form based for security architects to quantify usability in protocol design phase itself. The traditional tools, such as System Usability Scale (SUS), \cite{sus} normally do not take into account steps of the actual protocol development chain and can only be adopted after the core logic and UI prototype are designed. Also \cite{sus1} and \cite{design_with_usable_security} discussed that it is very rare to find expertise in security and usability in a typical developer. Building on our study of vulnerabilities and using base of findings from \cite{security_usability}, \cite{design_with_usable_security} and \cite{sus1}, we would like to reiterate that there is a need of integration of usability into design and protocol requirements so that security engineers and can take informed and usable decisions.


	
\begin{figure}
 		\centering
 			\subfigure[Needful adoptions]{	\includegraphics[width=.45\textwidth]{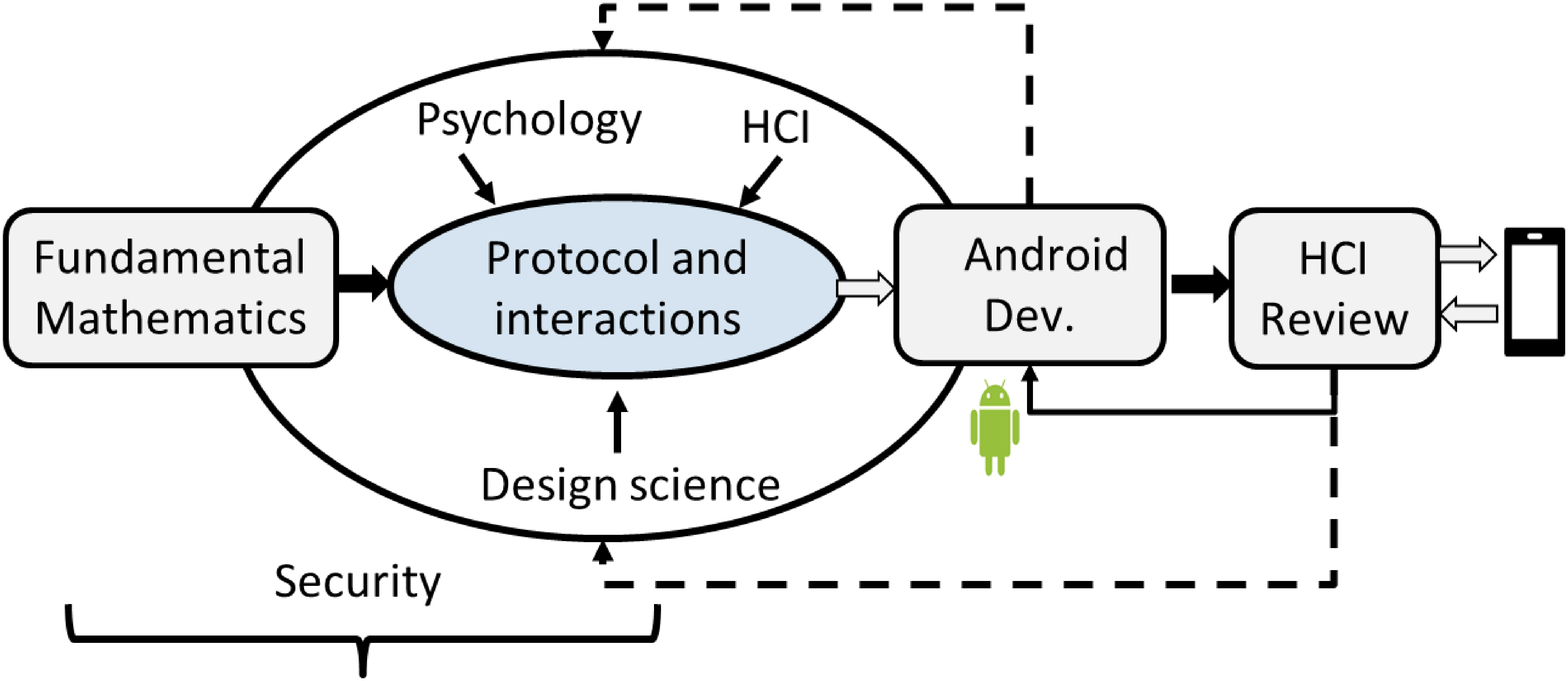}\label{changed}} \hfill
 		\subfigure[Lookup table, $T_{in}(U_{in})$]{	\includegraphics[width=.3\textwidth]{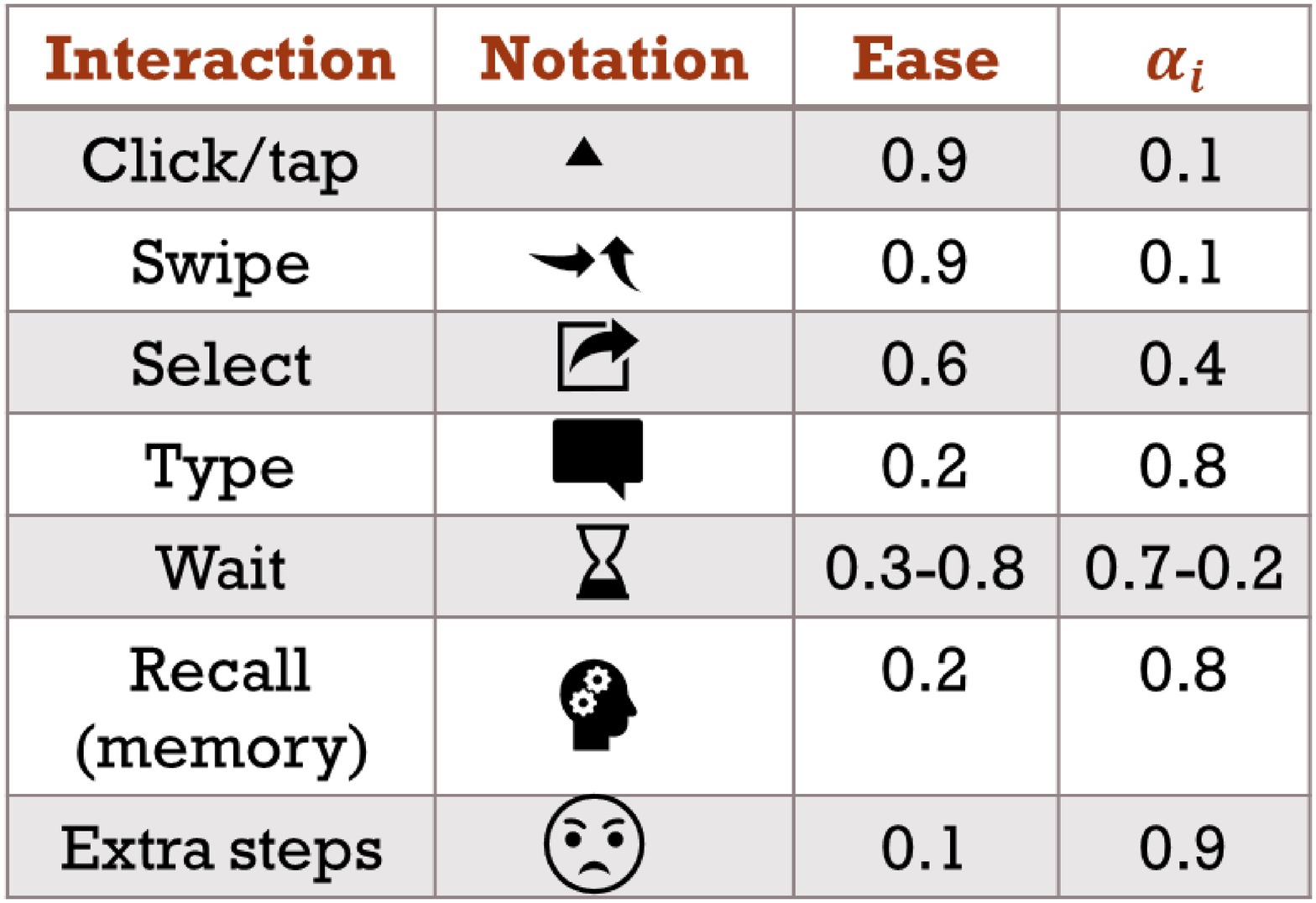}\label{table}}\vspace{-4mm}
 		\label{usability1}
		\caption{Knowledge transfer and lookup table for reference by protocol developers}
 	\end{figure}

	$\blacksquare$ \textbf{Approach:} Rather than depending primarily on usability studies and improvements, that too only after full-fledged design of applications, we argue that knowledge transformation and transfer would be a better strategy for usable security. As shown in Fig. (\ref{changed}), security protocol design can be thought of as two blocks: a) fundamental mathematics and b) protocol and interactions. 
	While designing steps of protocol and interaction points, experts can gain from collaborative attempts of researchers from heterogeneous domains including security, psychology, human-computer-interaction and design science and accommodate pre-defined suggestions. The system thus developed will have inherent usability and would require less time as well.
	

	


	
	
	\vspace{-8mm}
		\vspace{-6mm}

	$\blacksquare$ \textbf{Unification of System and User Space:}~ 
	The approach is inspired from \cite{sus1} which proposed a novel concept of Security Usability Symmetry (SUS) inspection method for usability measurement in early phase. In practice, a typical security system, product or service, can be jointly studied in two conceptual spaces: a) protocol space and b) user space. We modified UJM in attempt to bring these two spaces under one tool, set of guidelines, which helps in establishing a joint notion of security and usability quantification. We encourage protocol designers to consider the metrics as mentioned below:
\begin{enumerate}
	\item Define expected user steps at each block.
	\item Note down the points where user inputs or interactions, $U_{in}$, are expected in any form, categorize user interaction and assign a corresponding value, $\alpha_{i}$, from the lookup table, $T_{in}(U_{in})$, Fig (\ref{table}).
	\item Estimate the time taken, $t_{i+1}-t_i$, at every step of the system or protocol.
	\item Calculate the usability as reciprocal of user engagement, $\frac{1}{\sum_{i=1}^{N} \alpha_{i}(t_{i+1}-t_i)}$.
\end{enumerate}
This approach expects the determination of blocks, user interactions and time into computations to gauge usability. Based on our study, of ranking different modes of interactions, with 43 participating users, we provide a lookup table, Fig.(\ref{table}). It provides an example of interaction, $\alpha_{i}$, weights for typical actions that users encountered in smartphone applications. The clicks and taps, being the easiest of tasks, are lowest on interaction scores. The action which require recalling from memory needs additional cognitive effort for users, hence accounts to score of 0.8. Waiting for the UI to respond relates to the time. 



	

\vspace{-4mm}
\section{Related Work}\vspace{-2mm}






There are a number of works focus on automated mass analysis of sensitive method calls in Android applications (\cite{SCAnDroid} \cite{TaintDroid}).
Another technique for large-scale leak identification in Android applications is proposed in \cite{AndroidLeaks}. The authors utilise method mapping and taint analysis to reveal the privacy-sensitive functionality in an automated manner with a rate close to 800 APK per hour. Trade-offs between usability and security have been reported by several research works. \cite{good2003usability} studied KaZaA application from lenses of security and usability, and suggested that developers take too many assumptions regarding the users' knowledge of
file sharing, and violates secure interface guidelines. \cite{sus1} proposed a novel concept of Security Usability Symmetry (SUS) inspection method and the utilization of the Quality in Use Integrated Measurement Model (QUIM) for
model of usability measurement. Authors in \cite{survey}, \cite{survey2} and \cite{survey3} give details of security paradigms in D2D communication network which encompasses both in-band and out-band D2D pairing methods and cellular network facilitated exchanges under the framework of 3GPP LTE. A large portion of the works require Ad-hoc modes and support from network. Thus, \cite{d2d1} identified multiple attacks in Wi-Fi Direct-based D2D communications and introduced a short authentication-string-based key agreement protocol.
\vspace{-4mm}
	\section{Conclusion}\vspace{-2mm}
	We have studied the top D2D file sharing applications on Android which play a significant role in the offline sharing culture of their large userbase in India, China, Indonesia and a number of other fast-growing mobile markets, commonly referred to as Next Billion Users (NBU).
	In our analysis, we have identified a number of common insecure implementation flaws with an aim to understand the root causes behind them. Many of these flaws are caused by the usability requirements for the application flow and the limitations of its underlying protocols.
	We propose a methodology for early consideration of security risks through joint notion of the security and usability space.
	This view may help to identify and minimize usability bottlenecks in the system which motivate the security trade-offs in future implementations.

\bibliographystyle{splncs04}
\bibliography{bib}

\section{Appendix}

		\onecolumn
	\begin{longtable}{| p{.80\textwidth} | p{.30\textwidth} |p{.15\textwidth} |} \hline
	 	Improper username sanitization in ReceiverFragmentPeer.java in the \textbf{Google Files (com.google.android.apps.nbu.files)} through 1.0.220185905 allows the remote attacker to tamper with the Receiver`s connection confirmation & Reported to Google (Patched 08.02.2019)  \\ \hline
	 	The TCP communication turns into clear-text in the \textbf{Google Files (com.google.android.apps.nbu.files)} through 1.0.220185905 for Android if either the Sender or the Receiver uses Android 7.1.2, allowing an in-network attacker to sniff and tamper with Device-to-Device communication  & Reported to Google (Accepted)   \\ \hline
	 	Authentication token validation vulnerability in \textbf{Xender (cn.xender)} before 5.3.0.Prime allows attackers to remotely forge the write path and upload arbitrary files to the device filesystem. 
	     & CVE ID requested   \\ \hline
	 	A Path traversal vulnerability in static/storage/* in the \textbf{Xender (cn.xender)} before 4.8.0.Prime allows attackers to remotely retrieve arbitrary files from the device filesystem. 
		\textit{The vulnerability persists in the latest Xender 5.3.0.Prime.} & CVE ID requested Disclosed through Google Play Security Reward Program (Completed 20.12.2019)  \\ \hline
		A Path traversal vulnerability in waiter/downloadSharedFile in the \textbf{Xender (cn.xender)} before 4.2.2.Prime allows attackers to remotely retrieve arbitrary files from the device filesystem. 
		\textit{The vulnerability persists in the latest Xender 5.3.0.Prime.} & CVE-2018-19313 Disclosed through Google Play Security Reward Program (Completed 10.09.2019)  \\ \hline
		A reflected Cross-site scripting (XSS) vulnerability in the Web sharing functionality in the \textbf{SuperBeam (com.majedev.superbeam)} application through 4.1.3 for Android allows remote attackers to inject arbitrary JavaScript code via crafted URL to be executed on the client  & CVE-2018-19314   \\ \hline
		A Denial-of-Service (DoS) vulnerability in the \textbf{SuperBeam (com.majedev.superbeam)} application through 4.1.3  for Android allows attackers to drain the memory available to the application, resulting in a remote crash by scheduling a high number of invalid download requests  & CVE-2018-19315   \\ \hline
		In the \textbf{Superbeam (com.majedev.superbeam)} application through 4.1.3 for Android, the filenames of sent files are not sanitized and are rendered raw in the file list when received through the built-in web server endpoint on port 8080. The XSS, stored in the filename, is executed on the Receiver side.  & CVE-2018-19316   \\ \hline
		An insecure Wi-Fi access-point configuration in file-sharing functionality in the \textbf{SHAREit (com.lenovo.anyshare.gps)} application through 4.5.84 on Android 7.1, 7.1.1 and 7.1.2 allows the attackers to sniff and tamper with Device-to-Device communication & CVE-2018-19427   \\ \hline 
		An insecure Wi-Fi access-point configuration in the Send File functionality in the \textbf{Xender (cn.xender)} application through 4.2.2.Prime on Android 7.1, 7.1.1 and 7.1.2 allows attackers to sniff and tamper with Device-to-Device communication  & CVE-2018-19425   \\ \hline
		An insecure Wi-Fi access-point configuration in the Receive File functionality in \textbf{Zapya (com.dewmobile.kuaiya.play)} application through 5.7 (US) on Android 7.1, 7.1.1 and 7.1.2 allows attackers to sniff and tamper with Device-to-Device communication &  CVE-2018-19426  \\ \hline 
		An application package traversal vulnerability in the "Install SHAREit" widget served by a built-in web server in the \textbf{SHAREit (com.lenovo.anyshare.gps)} application through 4.5.84 for Android allows attackers to remotely enumerate installed application packages on the device and download them from device filesystem via apps/*.apk/?channel=webshare on TCP port 2999. 	\textit{The vulnerability persists in the latest SHAREit 5.0.88\_ww.}  & CVE-2018-19428 Disclosed through Google Play Security Reward Program (Completed 11.09.2019)   \\ \hline
		An insecure limitation of a Sender`s wireless network passphrase length, enforced by the Receiver user interface in the \textbf{Xender (cn.xender)} application through 4.2.2.Prime on Android facilitates remote attackers in password enumeration in order to associate with the device access point, sniff and tamper with device-to-device communication   & CVE-2018-19429   \\ \hline
		Cleartext file transmission via HTTP on port 6789 in WebShare mode in the \textbf{Xender (cn.xender)} application through 4.2.2.Prime for Android allows an in-network attacker to sniff and tamper with Device-to-Device communication  & CVE-2018-19430   \\ \hline
		Cleartext file transmission via HTTP on port 2999 in WebShare mode in the \textbf{SHAREit (com.lenovo.anyshare.gps)} application through 4.5.84 for Android allows an in-network attacker to sniff and tamper with Device-to-Device communication  & CVE-2018-19431   \\ \hline
		Anonymous FTP user, enabled by default in "Connect to computer" functionality in the \textbf{Xiaomi MiDrop (com.xiaomi.midrop) }application through 1.22.4 for Android allows an unauthenticated attacker to remotely download the entire storage of the Android device  & CVE-2018-19846   \\ \hline
		Unencrypted file transmission through FTP on port 2121 of the Android device in "Connect to computer" functionality in the \textbf{Xiaomi MiDrop (com.xiaomi.midrop)} application through 1.22.4 for Android allows the in-network attacker to sniff and tamper with files, transferred to and from the Android device & CVE-2018-19847   \\ \hline
		
			\caption{List of vulnerabilities, discovered during our analysis}
		\label{table:cve_list}
	\end{longtable}  \twocolumn{}
	\onecolumn

\end{document}